\documentclass{kapproc} 

\usepackage{epsf,amssymb}

\normallatexbib

\newcommand{\PT}{{\cal P}{\cal T}}
\newcommand{\CH}{{\cal H}}

\newcommand\hH{\CH}
\newcommand\eq{\begin{equation}}
\newcommand\en{\end{equation}}
\newcommand{\toline}[1]{}
\newcommand{\hepth}[1]{{\tt hep-th/#1}}
\newcommand{\condmat}[1]{{\tt cond-mat/#1}}
\newcommand{\quantph}[1]{{\tt quant-phys/#1}}
\newcommand{\nlinsys}[1]{{\tt nlin-si/#1}}
\newcommand\blank[1]{}

\newcommand\bea{\begin{eqnarray}}
\newcommand\eea{\end{eqnarray}}

\begin{document}

\articletitle[The ODE/IM correspondence and \dots]
{The ODE/IM correspondence and $\PT\!$-symmetric
quantum mechanics}

\author{\underline{Patrick Dorey}, Clare Dunning and Roberto Tateo}

\affil{
Department of Mathematical Sciences, University of Durham,\\
 Durham DH1 3LE, UK
(PED and RT)\\
Department of Mathematics, University of York, York YO1 5DD, UK
(TCD)}
\email{p.e.dorey@dur.ac.uk, roberto.tateo@dur.ac.uk, tcd1@york.ac.uk}

\vskip -6pt

\begin{abstract}
A connection between     
integrable quantum field theory and the spectral theory of
ordinary differential equations is reviewed, with particular
emphasis being given to its relevance to certain problems in
$\PT\!$-symmetric quantum mechanics.
\\
(Contribution to the proceedings of the NATO Advanced Research Workshop
on Statistical Field Theories, Como, 18-23 June 2001.)
\end{abstract}

\vskip -6pt


\section{Introduction}

This talk concerned a recently-discovered relationship between a particular
class of  integrable models and the spectral theory of 
ordinary differential equations in complex domain, a link which is
sometimes referred to as the `ODE/IM correspondence'.
We recently reviewed many aspects of this in \cite{DDTrev} (see also
\cite{DDTrevb}), and a more
extended version of \cite{DDTrev} is currently in preparation.
Therefore in this contribution we shall content ourselves with a quick
sketch of some of the main features of the story, 
updating \cite{DDTrev}
as we go along with some references to more recent developments.

Two previously-distinct areas of investigation form the backdrop to
this work. On the one hand, there is the theory of integrable lattice
models and integrable quantum field theories, in particular as
extensively
developed by Baxter~\cite{Bax} and then Kl\"umper, Pearce and
collaborators \cite{KP,KBP} on the lattice side, and 
by Bazhanov, Lukyanov and
Zamolodchikov \cite{BLZ1,BLZ2,BLZ3}
on the integrable quantum field theory side.
On the other, there is an approach to the theory of ordinary
differential equations via functional relations, pioneered by Sibuya
\cite{Sha} and Voros \cite{Voros}.
The first observation of a concrete connection between these two
subjects was made in \cite{DTa}; 
related subsequent work includes
\cite{BLZa, Sa, FAiry, DTb, DTc, Sb,Sc,DDTa,Srev,Hikami,DDTb,DDTc,BHK}. 
Recently, the relevance of the ODE/IM correspondence to certain
problems in
$\PT\!$-symmetric quantum mechanics, first observed in \cite{DTb},
was re-emphasised \cite{DDTb,DDTc}; since this also formed the 
central theme of the
talk given at the conference, the next section is devoted to
an introduction to this topic.

\section{${\cal PT}\!$-symmetric quantum mechanics}
Some years ago,
considerations of the
Yang-Lee edge 
singularity in two dimensions
led Bessis and Zinn-Justin \cite{BZJ} to ask themselves about the spectrum of
the following `quantum-mechanical' Hamiltonian:
\eq
\hH=p^2+ix^3\,.
\label{proba}
\en
This is a cubic oscillator, with purely imaginary coupling $i$. 
To make the question precise, we
shall say that
the (possibly complex) number $E$ is in the spectrum of
(\ref{proba}) if and only
if
the equation $\hH \psi(x)=E \psi(x)$ has a 
solution 
$\psi(x)$ 
lying in $L^2(\mathbb R)$\,.
For the Hamiltonian (\ref{proba}), this 
is equivalent to demanding that $\psi(x)$ should decay as
$x\to\pm\infty$ along the real axis:
\[
\epsfxsize=.5\linewidth
\epsfbox{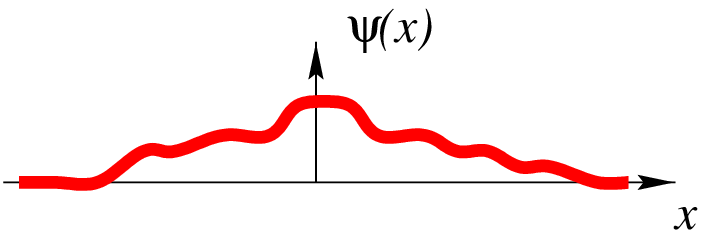}
\] 
Even for real $x$
and $E$\,, the wavefunction $\psi$ cannot be everywhere real.
Furthermore, the usual
argument leading to reality of the eigenvalues does not apply here, as
the Hamiltonian is not Hermitian. Nevertheless,  perturbative and
numerical studies led 
Bessis and Zinn-Justin to the following claim:

\smallskip

\noindent
{\bf Conjecture 1} \cite{BZJ}:
the spectrum of $\hH$ is real and positive.

\smallskip

\noindent

\begin{figure}[ht]
\centerline{\epsfxsize=.7\linewidth\epsfysize=.64\linewidth%
\epsfbox{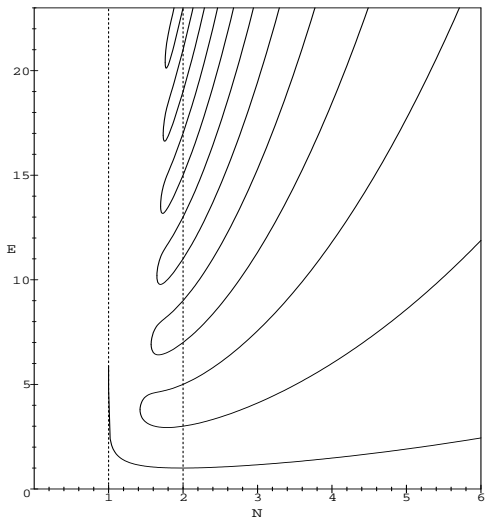}}
\vspace{-1.2cm}
\vspace{.1cm}
\caption{Eigenvalues of 
$\CH_N\psi=E\psi$\,,
plotted as a function of $N$.}
\label{fig2}
\end{figure}

In 1997, Bender and Boettcher~\cite{BB} suggested an interesting
generalisation. Suppose $N$ is a positive real
number, and consider the spectrum of the Hamiltonian
\eq
\hH_N=p^2-(ix)^N\qquad\quad\mbox{($N$ real, $>0$)}
\label{probb}
\en
where $\psi(x)$ is again required to lie in $L^2({\mathbb R})$.
For $N=3$, this is the Bessis-Zinn-Justin problem (\ref{proba}),
while for $N=2$ it reduces to the much better-understood
simple Harmonic oscillator. 
For non-integer values of $N$, the
`potential' $-(ix)^N$ is not single-valued, so a branch cut should
be added running up the positive $x$-axis.
Once this has been done, the problem as stated is well-defined (at
least for $N<4$) and can be studied numerically. This is what Bender
and Boettcher did, with a surprising result:
while the spectrum for $N\ge 2$ is real, as $N$ decreases below $2$,
infinitely-many eigenvalues pair off and become complex,
with only finitely-many remaining real. By the time $N$ reaches $1.5$,
all but three have become complex, and as $N$ tends to
$1$ the last real
eigenvalue diverges to infinity. 
This curious behaviour
is shown in figure~\ref{fig2}, taken from \cite{DTb}; it
reproduces the results of \cite{BB}.

For $N\ge 2$, the fact that their numerically-obtained spectrum was 
real and positive led Bender and Boettcher to generalise
the Bessis-Zinn-Justin
conjecture to

\smallskip
\noindent
{\bf Conjecture 2} \cite{BB}:
the spectrum of $\hH_N$ is real and positive for $N\ge 2$.\\

At generic values of $N$, (\ref{probb}) is no more Hermitian than
(\ref{proba}), but it does have a property, emphasised by Bender and
Boettcher, known as
$\PT$ symmetry. 
(\,`${\cal P}$', or 
parity, acts by sending $x$ to $-x^*$ and $p$ to $-p$ while ${\cal T}$,
time reversal, 
sends $x$ to $x$, $p$ to $-p$ and $i$ to $-i$, so the combined effect
of $\PT$ on a potential $V(x)$ is to send it to $V(-x^*)^*$\,.\,)
This symmetry does not by itself 
guarantee reality of the spectrum, as is clearly demonstrated by
figure \ref{fig2}, but it does imply that the 
eigenvalues are either
real, or appear in complex-conjugate pairs. (This is analogous to 
the behaviour of the roots of a real polynomial.)
In spite of this fact,
reality proofs in $\PT\!$-symmetric quantum mechanics have 
been surprisingly elusive, and prior to the recent application of
the ODE/IM correspondence to the problem \cite{DDTb}, even
the simple-to-state conjectures 1 and 2 were unproven:
see \cite{BBN,DP,DT,Mez,Shin,BW,Mez1} for some
earlier discussions of the issues involved.

\begin{figure}[ht]
\centerline{\epsfysize=.36\linewidth
\epsfbox{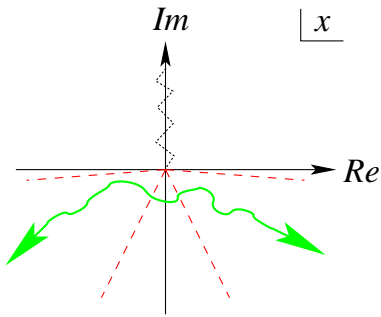}}
\caption{
A possible wavefunction contour for $N>4$.
}
\label{sectors}
\end{figure}

One further detail concerning figure \ref{fig2} needs care: 
when $N$ hits
$4$, the naive definition of the eigenvalue problem runs
into difficulties. 
This can be traced to the fact that the asymptotic behaviour of the
wavefunction along the real axis becomes oscillatory at $N=4$.
To avoid this difficulty, 
the contour along
which $\psi(x)$ is defined should be deformed down from the real
axis into the complex plane, so as to continue to avoid the so-called
anti-Stokes lines for the problem. This point is explained at greater
length in \cite{BB,DDTrev}, and so we shall not linger on it here.
However, in figure \ref{sectors} we show one possible contour for $N$
just larger than $4$.

Finally, in \cite{DTb} we suggested a further generalisation 
of the Bessis-Zinn-Justin conjecture, again $\PT\!$-symmetric,
by including an
angular-momentum term $l(l{+}1)x^{-2}$\,:
\eq
\hH_{N,l}=p^2- (ix)^{N} + l(l{+}1)/x^2.
\label{leqn}
\en
Numerical work gave us strong evidence for 
\smallskip

\noindent
{\bf Conjecture 3} \cite{DTb}:
the spectrum of $\hH_{N,l}$ is real and positive for 
$N\ge 2$
and 
$l$ small.
\medskip

{~}
\begin{figure}[ht]
\vspace{-0.6cm}
\[\begin{array}{ll}
\epsfxsize=.43\linewidth\epsfysize=.43\linewidth\epsfbox{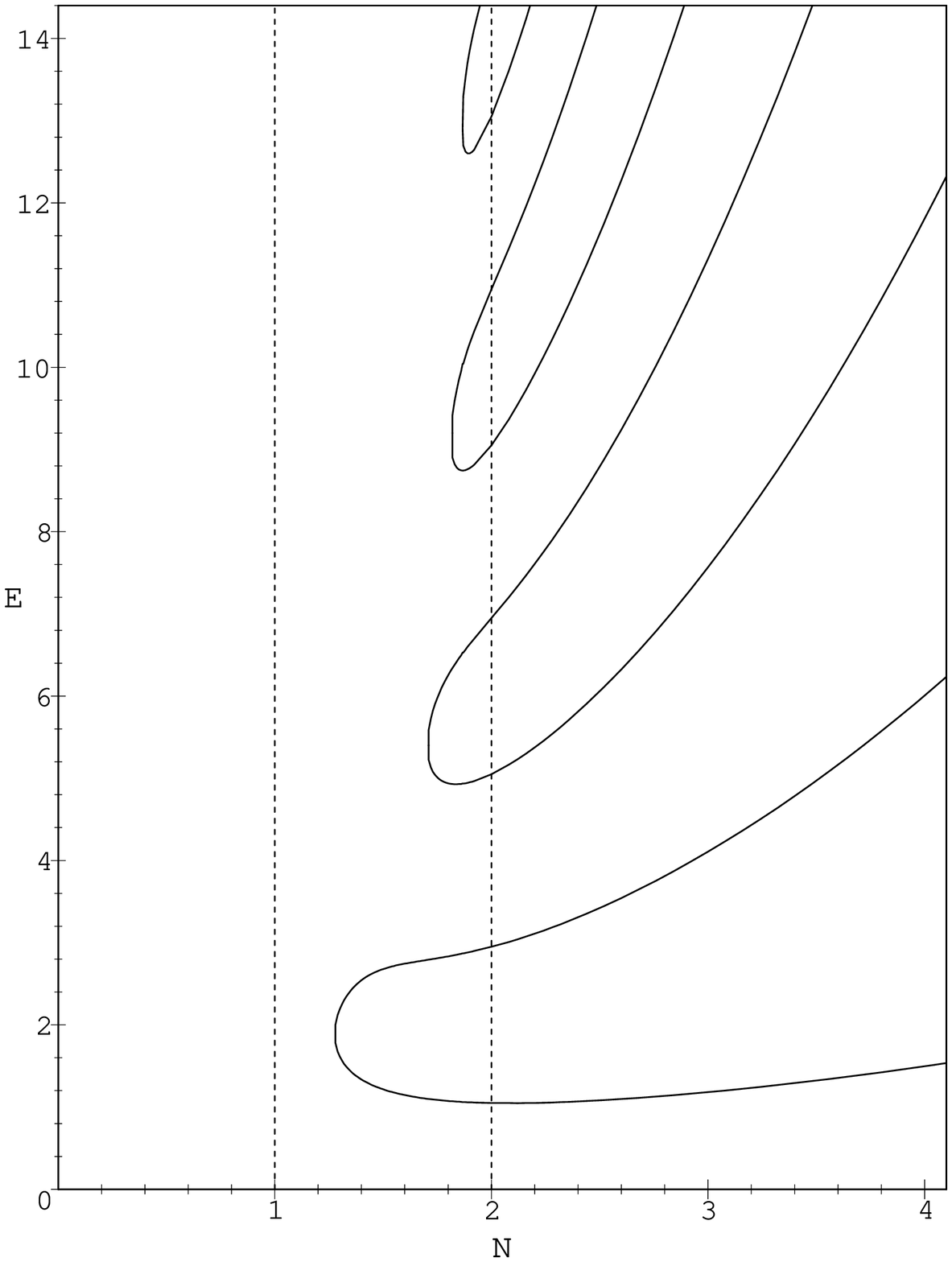}
{}&
\epsfxsize=.43\linewidth\epsfysize=.43\linewidth\epsfbox{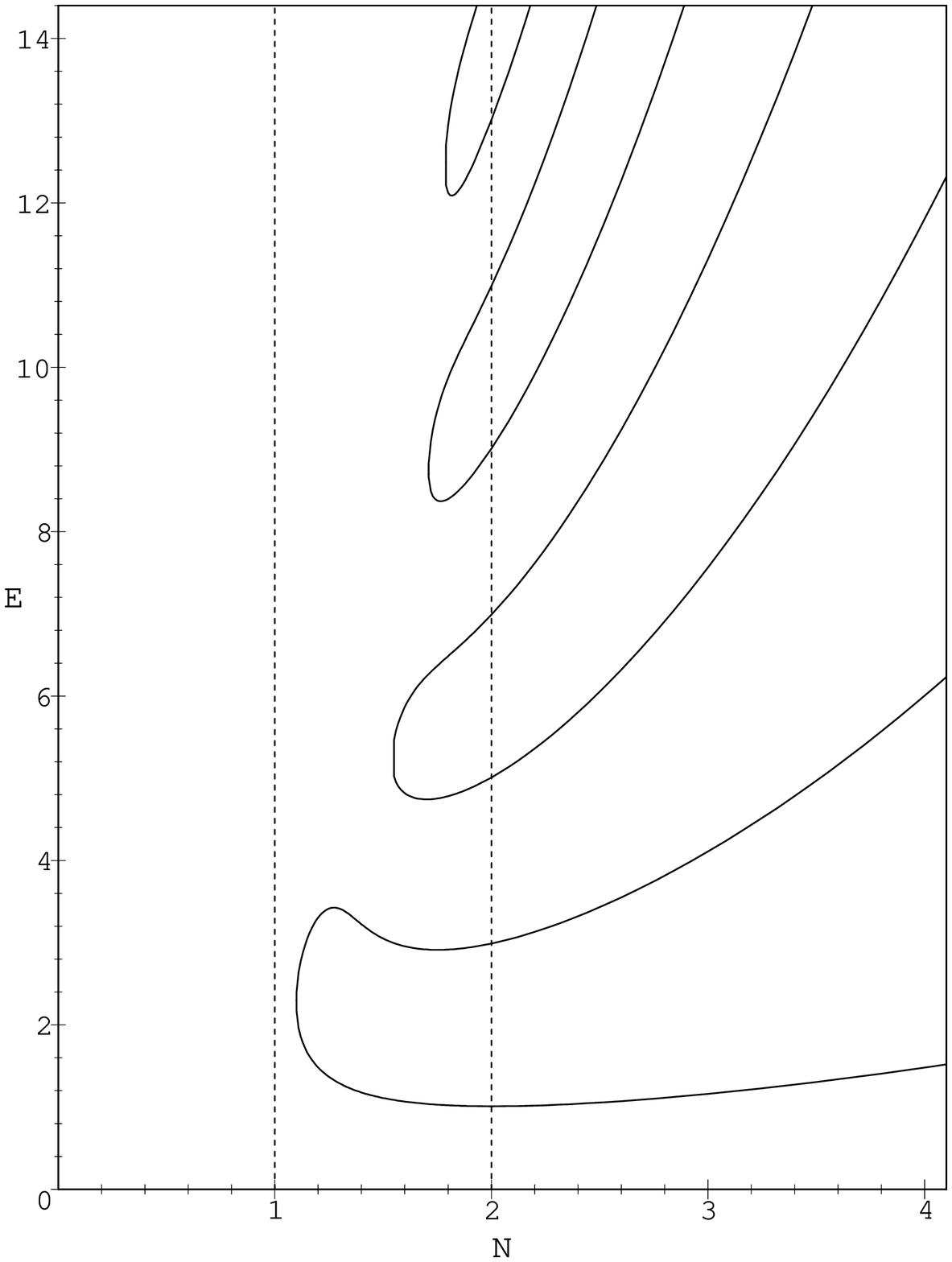}
\\[-15pt]
\parbox[t]{.4\linewidth}{\quad~~~\small 3a) $l=-0.025$}
{}~&~
\parbox[t]{.4\linewidth}{\quad~~\small 3b) $l=-0.005$}
\\[-7pt]
\epsfxsize=.43\linewidth\epsfysize=.43\linewidth\epsfbox{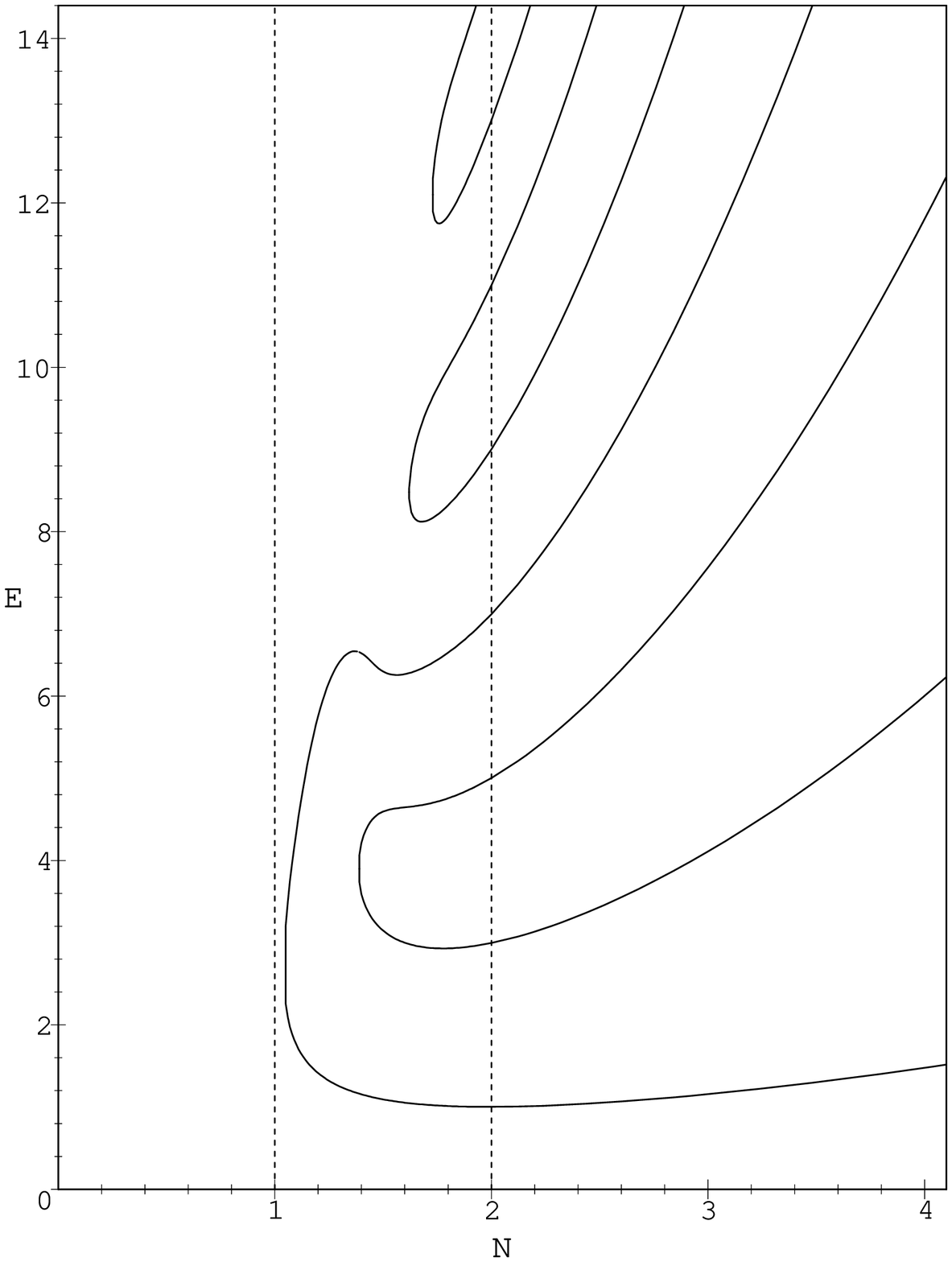}
{}&
\epsfxsize=.43\linewidth\epsfysize=.43\linewidth\epsfbox{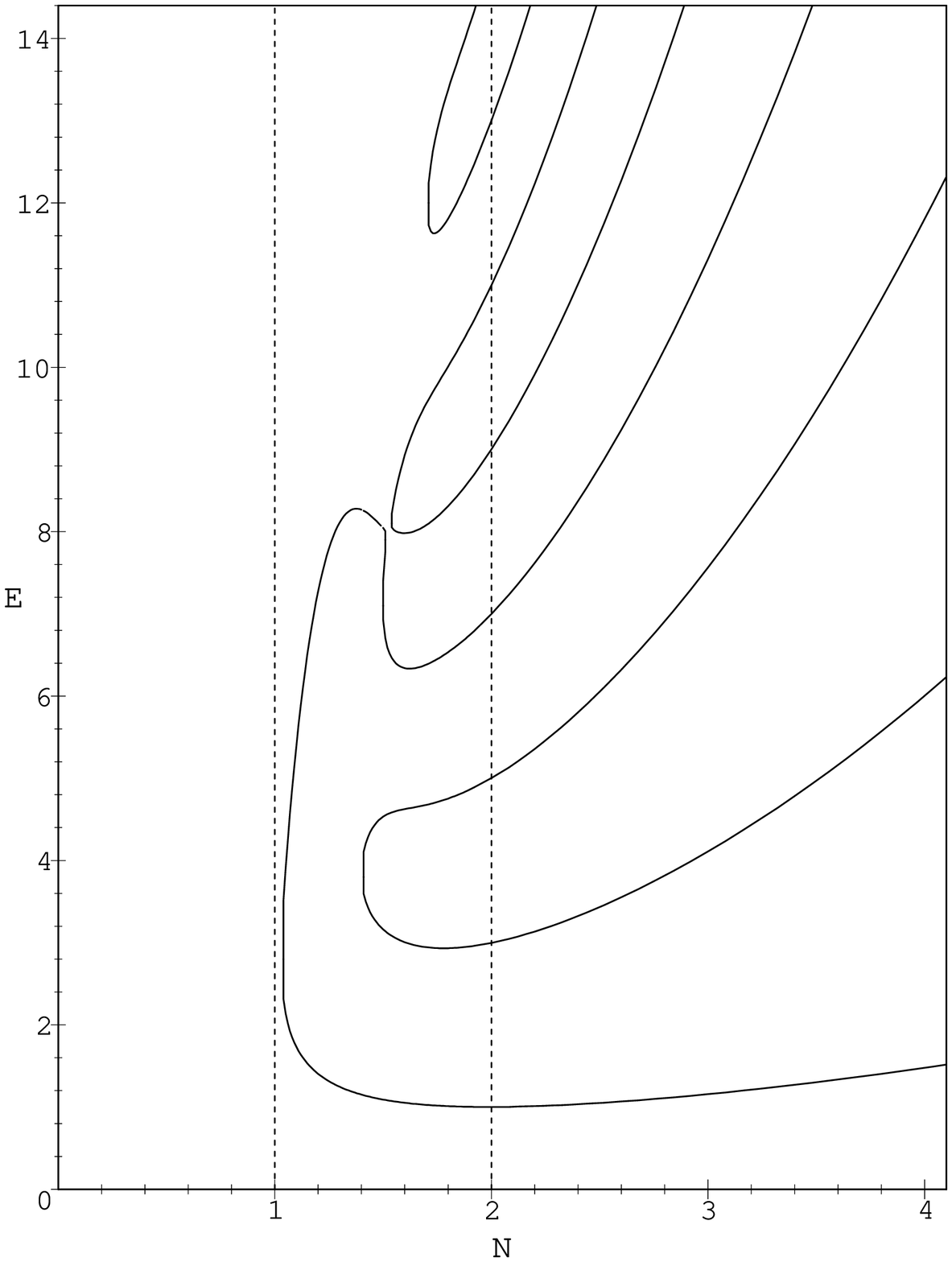}
\\[-15pt]
\parbox[t]{.4\linewidth}{\quad~~~\small 3c) $l=-0.0015$}
{}~&~
\parbox[t]{.4\linewidth}{\quad~~\small 3d) $l=-0.001$}
\\[-7pt]
\epsfxsize=.43\linewidth\epsfysize=.43\linewidth\epsfbox{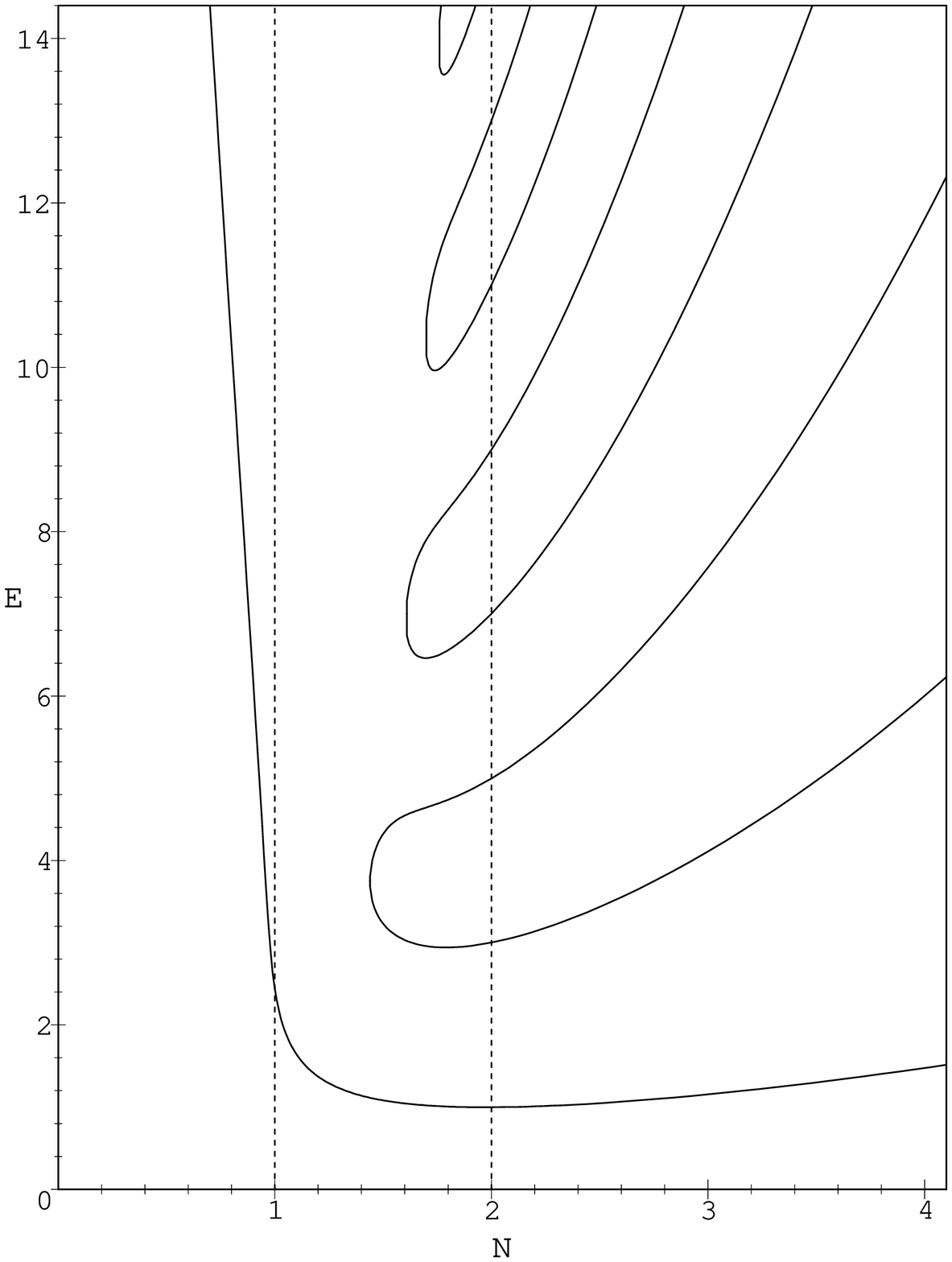}
{}&
\epsfxsize=.43\linewidth\epsfysize=.43\linewidth\epsfbox{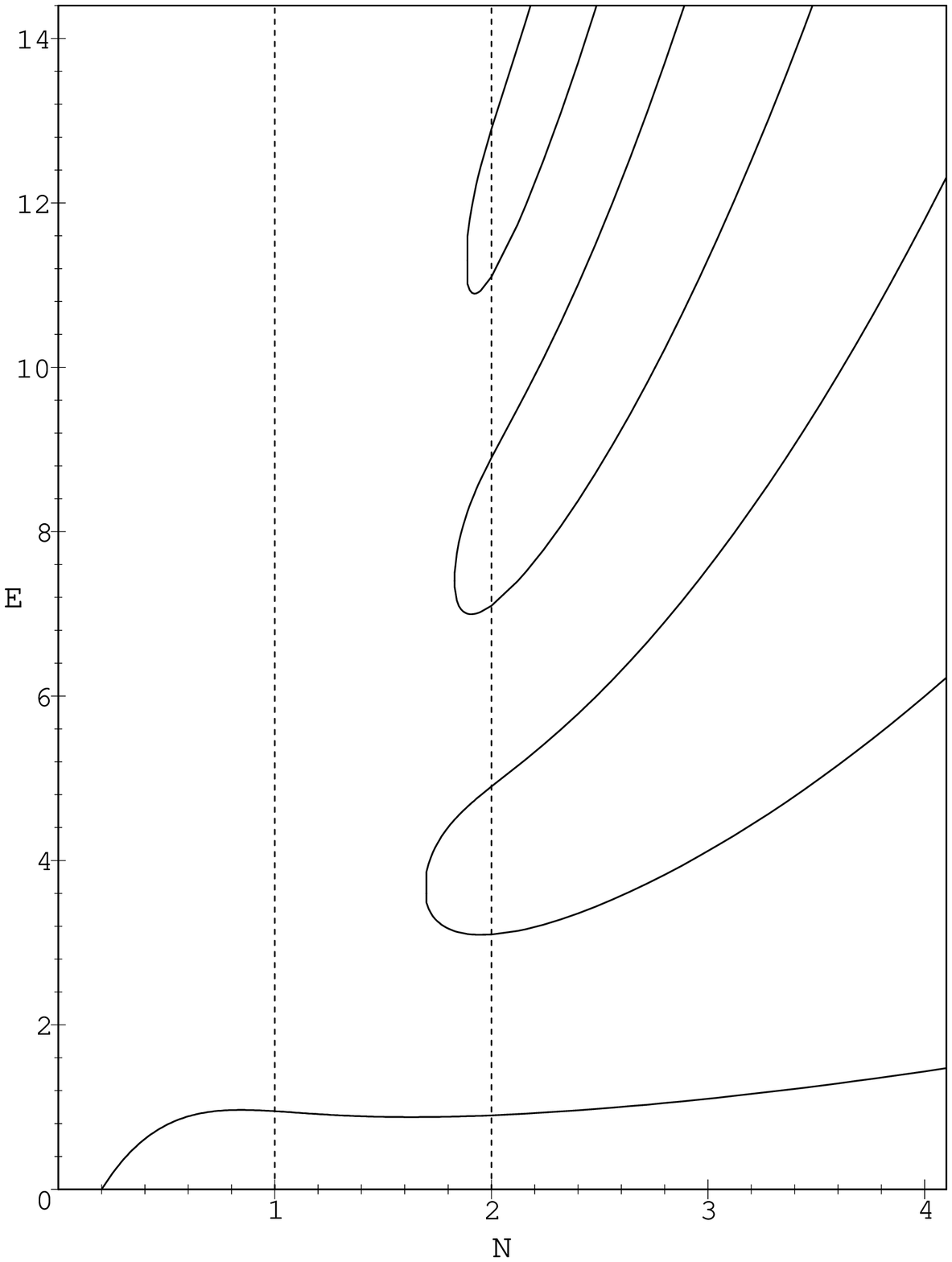}
\\[-15pt]
\parbox[t]{.4\linewidth}{\quad~~~\small 3e) $l=0.001$}
{}~&~
\parbox[t]{.4\linewidth}{\quad~~\small 3f) $l=0.05$}
\\
\end{array}\]
\vskip -0.3cm
\caption{Eigenvalues of 
$\hH_{N,l}\psi=E\psi$\,,
plotted as a function of $N$ at various nonzero values of $l$.}
\label{fig3}
\vskip -0.3cm
\end{figure}
While the angular-momentum term does not affect
the reality of the eigenvalues for $N\ge 2$, it can make
a remarkable difference
to the way in which they become complex.
Figure \ref{fig3} shows a sequence of spectral plots for $l$ varying
from $-0.025$ to $0.05$.
Notice that for $l=-0.025$, the
connectivity of the real eigenvalues is
completely
reversed from that seen in figure \ref{fig2}, so that while for
$l{=}0$ the first and second
excited states pair off, at $l=-0.025$ 
the first excited state
is instead paired with 
the ground state, and so on up the spectrum. 
With this in mind it might be hard to see how
it is possible to pass between the two sets of spectra 
simply by varying the continuous parameter $l$ 
from $-0.025$ to zero. 
The intermediate plots shown in figures~\ref{fig3}b--\ref{fig3}d 
should help the reader to resolve
this question, while figures \ref{fig3}e and \ref{fig3}f show how
a real eigenvalue ventures into the region $N<1$ once $l$ becomes
positive.

We hope that this discussion will have convinced the reader of the
surprising richness of $\PT\!$-symmetric quantum mechanics. The field
has now developed a considerable momentum of its own right, with many
other examples being discussed in the literature -- 
refs.~\cite{Za,BCQ,HX,BBJMS,CIRR,BS,TF,ZJ1} 
offer a
selection of this work. The conceptual basis of the subject is also
a topic of interest, with recent discussions to be found in
\cite{ZJ0,BQZ,ZJ2,GS,KS,AM}. 
The ODE/IM correspondence has little to say directly about
this question, but it has yielded what to our knowledge is the first complete
proof of spectral reality for a family of non-trivial $\PT\!$-symmetric 
problems, including conjectures 1--3 above as special 
cases~\cite{DDTb}. 

Limitations of space preclude a complete explanation of how the proof
goes, but in the remainder of this article we shall at least describe
the basic elements of the correspondence which led to it.

\section{Functional relations in integrable models}
\label{funim}
It was Baxter who first observed that an alternative to the Bethe
ansatz method of solution for the six-vertex model was provided by a
functional equation, now called a T-Q relation \cite{Bax}. 
Related ideas were then used with much success in other lattice
models (see for example \cite{KP,KBP}), but it was not until the work
of Bazhanov, Lukyanov and Zamolodchikov that they were systematically
applied to integrable quantum field theories defined directly in the
continuum \cite{BLZ1,BLZ2,BLZ3} (elements of this structure had also
been observed by Fendley, Lesage and Saleur~\cite{FLS}).
A key feature of the work~\cite{BLZ1,BLZ2} was 
the construction of two continuous families of commuting operators
acting in the Hilbert space of the quantum field theory, defined on a
circle:
${\bf T}(\lambda)$, sometimes called the quantum
transfer matrix, and  ${\bf Q}(\lambda)$.
These were shown to satisfy the following relation, the field-theory
analogue of Baxter's T-Q relation:
\eq
{\bf T}(\lambda){\bf  Q}(\lambda)=
{\bf Q}(q^{-1}\lambda)+{\bf Q}(q\lambda)\,,
\en
where $q=e^{i\pi\beta^2}$, and $\beta$ is a parameter characterising
the particular integrable field theory under consideration.
Each theory has a set of ground states (`$\theta$-vacua')
$|p\rangle$. These are labelled by a `momentum' $p$ ({\em not} the
same as the momentum operator of the last section)
and are
eigenstates
of the ${\bf T}(\lambda)$ and ${\bf Q}(\lambda)$ operators.
If we focus on the corresponding eigenvalues by setting
$T(\lambda,p)=\langle p|{\bf T}(\lambda)|p\rangle$,
$A(\lambda,p)=\lambda^{2 p/ \beta^2}\langle p|{\bf
Q}(\lambda)|p\rangle$, then the
T-Q relation for these vacuum eigenvalues can be written as
\eq
T(\lambda,p)A (\lambda,p)= 
e^{- 2\pi ip}A (q^{-1}\lambda,p)
+ e^{+ 2\pi ip}A (q\lambda,p)\,,
\label{tqeq}
\en
an equation that was also obtained in~\cite{FLS}.
The prefactor inserted into the definition of 
$A(\lambda,p)$ ensures that it, like $T(\lambda,p)$, is
a single-valued function of $\lambda^2$.
But not only are $A$ and $T$ single-valued, they are also {\em entire}
functions of $\lambda^2$.
This makes (\ref{tqeq}) a very powerful
constraint: 
when combined with the leading asymptotics of $T$ and
$A$, it admits just a discrete, albeit infinite, set of solutions
\cite{Bax,BLZ2}.
The ground state eigenvalues are known to have further analyticity
properties, connected with the distribution of their zeroes,
which allow even this ambiguity to be removed.
{}From the point of view of integrable models, all of this is important
because the eigenvalues $T(\lambda)$ and $A(\lambda)$ encode the
values of an infinite set of conserved quantities acting on the
corresponding states; for the ODE/IM correspondence, the immediate
relevance is rather that it allows a precise link to be established
with certain quantities arising in the study of ordinary differential
equations, as will now be described.

\section{Functional relations in differential equations}
\label{funode}

Surprisingly, the T-Q relation  introduced  in the last section
also governs the problems in $\PT\!$-symmetric quantum mechanics
discussed in section 2.
For convenience we shift $x$ to $x/i$ and $E$ to $-E$
so that the ODE associated with
(\ref{leqn}) is
\eq
\left[-\frac{d^2}{dx^2}+ x^{N} + \frac{l(l{+}1)}{x^2}-E\right]
\psi(x)= 0~.
\label{sibeq}
\en
Considering this equation
for  general complex values of $x$, it is a relatively simple matter
to establish the 
following Stokes relation:  
\eq
C(E,l)y(x,E,l)=
\omega^{-1/2}y(\omega x,\omega^{-2}E,l)+
\omega^{1/2}y(\omega^{-1}x,\omega^2E,l)~,
\label{tqyrelx}
\en
where $\omega=e^{2\pi i/(N{+}2)}$, and
$y(x,E,l)$ is a particular solution to (\ref{sibeq})  
vanishing as $x \rightarrow \infty$ along the real axis
and  uniquely determined 
by its asymptotic behaviour there (see \cite{DTb} for more details).
{}From another perspective,
any solution to (\ref{sibeq}) 
can be written as a linear combination of a solution, $\psi^+(x)$,
behaving near $x=0$ as $x^{l+1}$, and
one, $\psi^-(x)$, behaving there as $x^{-l}$. 
Setting $y(x,E,l)=D(E,l) \psi^-(x) + D(E,-1{-}l) \psi^+(x)$, 
(\ref{tqyrelx}) implies that
\eq
C(E,l)D(E,l)=
\omega^{-(l+1/2)}D(\omega^{-2}E,l)+
\omega^{(l+1/2)}D(\omega^2E,l)~.
\label{cdeq}
\en
The functions $D(E,l)$ and $C(E,l)$
are entire in $E$, 
and have analogous analyticity properties to the ground-state
eigenvalues $T(\lambda)$ and $A(\lambda)$ described in the last
section. Together with the
obvious similarity of form between (\ref{tqeq}) and (\ref{cdeq}), this
permits a precise relationship to be established between the two sets
of objects, and this was the approach to the ODE/IM correspondence
that was adopted in \cite{DTb}. (It is also possible to proceed via
so-called `quantum Wronskian' relations, as in~\cite{DTa,BLZa}.)

{}From its definition, the  zeroes of $D(E,l)$ are
the values of $E$ at which, for $l>-1/2$, the function $x^{-1/2} y(x)$ 
vanishes both
at $x=\infty$ and $x=0$.
This means that $D$ can be interpreted as a spectral determinant, for
a problem which, in contrast to the problems of
section 2, {\em is} Hermitian. 
The $\PT\!$-symmetric problems, on the other hand, turn out to be
encoded in the zeroes of $C$ (or equivalently $T$)\,:
the Stokes multiplier  $C(-E,l)$  vanishes if and only if, at that
value of
$E$, 
(\ref{leqn}) has a nontrivial solution which decays to
zero as $x$ tends to infinity
on the quantisation contour  shown in figure~\ref{sectors}.
As a result, $C$ is
a spectral determinant for the Bessis-Zinn-Justin-Bender-Boettcher
problem,
and its generalisation~\cite{DTb}
to non-zero angular
momentum. Even better, the
zeroes of  $C$ are constrained by the T-Q relation.  By
combining this with the positive-reality of the zeroes of $D$ -- 
reality being a consequence
of the Hermiticity of the problem that $D$ encodes --
it is not too hard, for $N>2$,
to prove that the zeroes of $C$ must be real,
thus settling conjectures 1--3 of section~2~\cite{DDTb}.
In fact, in \cite{DDTb} we were able to prove a slightly more general
result, but we refer the reader to that paper and
to \cite{DDTc} for details.

\section{Conclusions}
\label{concl}
A principal aim of this talk has been to give some hints of
the link between two fascinating
research fields: the spectral theory
of ordinary differential  equations, and
the theory of integrable models.
Whether results from ordinary differential equations are used in the
study of integrable models, or vice versa,
is largely a matter of taste and personal background;
in either case the
correspondence promises to be  very  fruitful.
Besides the proof of the reality conjectures mentioned above,
the non-linear integral equation technique, an  established  tool in
integrable models,
has been successfully  introduced into the study of
spectral problems \cite{DTa,DTb,Sc} (figures \ref{fig2} and \ref{fig3} are
concrete examples of
this application),
 and  Bethe ansatz equations have been  used
to derive spectral equivalences~\cite{DDTb}.
Surprisingly, the  latter result links  integrable models to the
recent and exciting  discovery of  polynomial generalisations of
supersymmetry
in quantum mechanics \cite{Adr,KP0,KPl,Ao0,Ao1,RT}.
In the other direction,
certain duality relations, important to the
condensed matter physics applications of integrable models,
can be now   proved  by   simple variable changes in the relevant
differential equations \cite{BLZa,DTc,DDTa,BHK}.

\begin{acknowledgments}
{
\small
\noindent
PED thanks
the organisers 
for the invitation to speak
at the conference;
TCD and RT thank the UK EPSRC for a Research Fellowship
and an Advanced Fellowship respectively. RT was also
supported by an EPSRC VF grant, number~GR/N27330.
}
\end{acknowledgments}

\section*{Note added}
Ref.~\cite{Shin2} appeared as
we were preparing a version of this article for submission to the
electronic archives. In this very interesting preprint, Shin observes
that positive-reality of the zeroes of $D(E,l)$ is not a necessary condition
for the reality proof discussed at the end of section~4 above to go 
through -- it can be weakened considerably. This allows the argument to
be generalised to 
cover a greatly expanded class of potentials, and shows that the
relevance of ideas inspired by the ODE/IM correspondence to
problems in $\PT\!$-symmetric quantum mechanics goes much further than
had previously been suspected.

\end{document}